\documentclass[useAMS,usenatbib,longnamesfirst,usedcolumn]{mn2e} 
\usepackage{epsfig}	
\usepackage{times}	
\usepackage{subfigure}	
\title[Abundance profiles of groups]{Abundance profiles and cool cores in galaxy groups}
\author[R. Johnson, A. Finoguenov, T.J. Ponman, J. Rasmussen \& A. J. R. Sanderson]
       {Ria Johnson$^{1}$\thanks{Contact via: tjp@star.sr.bham.ac.uk},
        Alexis Finoguenov$^{2}$, Trevor J. Ponman$^{1}$, 
        Jesper Rasmussen$^{3}$, \newauthor  Alastair J. R. Sanderson$^{1}$\\
        $^{1}$School of Physics and Astronomy, University of
        Birmingham, Edgbaston, Birmingham B15 2TT, UK\\
        $^{2}$Max-Planck-Institut f\"{u}r extraterrestrische Physik,
        Giessenbachstra{\ss}e, 85748 Garching, Germany\\
	$^{3}$Dark Cosmology Centre, Niels Bohr Institute, University of Copenhagen, DK-2100 Copenhagen, Denmark
         }

\pagerange{\pageref{firstpage}--\pageref{lastpage}}
\pubyear{2010}

\shortcites{RC3,degrandi04,peres98,voit03b,paturel03,napolitano05,mahdavi05,finoguenov06,finoguenov07,baldi07,sun08,sivanandam08,kochanek01,mccarthy08,arnaud92,buote03b,rebusco05,rasera08,jetha07,rebusco06,moll07,burns08,poole08,best07,shabala08,allen01b,gu07,condon98,coziol09,sato09,tokoi08,morita06,komiyama09,kirkpatrick09}

\newcommand{\ASCA}{\emph{ASCA}}
\newcommand{\Chandra}{\emph{Chandra}}
\newcommand{\Rosat}{\emph{ROSAT}}
\newcommand{\BSAX}{\textit{BeppoSAX}}
\newcommand{\keV}{\ensuremath{\mbox{~keV}}}
\newcommand{\kpc}{\ensuremath{\mbox{~kpc}}}
\newcommand{\LR}{\ensuremath{L_{\mathrm{1.4GHz}}}}

\newcommand{\XMM}{\emph{XMM-Newton}}
\newcommand{\RF}{\ensuremath{r_{500}}}

\newcommand{\rr}{\textsc{r}}
\newcommand{\Zsol}{\ensuremath{Z_{\odot}}}
\begin{document}

\maketitle

\label{firstpage}

\begin{abstract}
\noindent 
Using data from the Two Dimensional \XMM\ Group Survey
(2dXGS), we have examined the abundance profile properties of both
cool core (CC) and non cool core (NCC) galaxy groups. The ten NCC
systems in our sample represent a population which to date has been
poorly studied in the group regime. Fitting the abundance profiles as
a linear function of log radius, we find steep abundance gradients in
cool core (CC) systems, with a slope of $-0.54\pm0.07$.
In contrast, non cool core (NCC) groups have profiles consistent with 
uniform metallicity. Many CC groups show a central abundance dip or plateau,
and we find evidence for anticorrelation between the
core abundance gradient and the 1.4\,GHz radio power of the brightest
group galaxy (BGG) in CC systems.
This may indicate the effect of AGN-driven mixing
within the central $\sim$0.1\,\RF. It is not possible to
discern whether such behaviour is present in the NCC groups, due to the small
and diverse sample with the requisite radio data. The lack of strong
abundance gradients in NCC groups, coupled with their lack of cool
core, and evidence for enhanced substructure, leads us to favour
merging as the mechanism for disrupting cool cores, although we cannot
rule out disruption by a major AGN outburst. Given the implied
timescales, the disruptive event must have occurred within the past
few Gyrs in most NCC groups.
\end{abstract}
\section{Introduction}
\label{sec:intro}
The dominant baryonic mass component in galaxy groups and clusters is
the hot X-ray emitting intracluster medium (ICM), with stellar mass
becoming increasingly important as mass decreases
\citep{gonzalez07,giodini09}. Studying the metallicity of the ICM can
provide insight into the processes that have shaped its thermodynamic
history. The heavy elements observed in the ICM originate
predominantly from supernovae explosions, which eject material into
the ICM \citep{arnaud92}. Gas can also be removed from galaxies, thus
enriching the ICM with metals, through processes such as ram-pressure
stripping \citep{gunn72} and galaxy-galaxy interactions
\citep[see][for a review of enrichment processes]{schindler08}. The
efficiency of any one transport process depends on the properties of
both the galaxies and their large-scale environment \citep[][and
references therein]{schindler08}.

Using \ASCA\ and \Rosat\ observations, \citet{finoguenov99} found the
groups HCG\,62 and NGC\,5044 to have significant negative abundance
gradients, a result also seen in the \Rosat\ sample of
\citet{buote00}. The abundance gradient in NGC\,5044 was also observed
by \citet{buote03b} using \XMM\ and \Chandra\ data, and recent studies
have shown the presence of an abundance gradient to be a common
feature
\citep[e.g.][]{morita06,rasmussen07,tokoi08,komiyama09,sato09}.
\citet{rasmussen09} also find a central excess of iron in all but two
of their groups, the presence of which can be explained solely by
supernovae Type-Ia products from the central galaxy. The excess
extends beyond the optical limits of the central galaxy, as also seen
in clusters \citep{david08,rasera08}. The re-distribution of enriched
gas can be achieved by outflows from active galactic nuclei (AGN)
\citep[e.g.][]{mathews04,rebusco06,moll07}, the presence of which is
commonly invoked to explain the lack of catastrophic cooling in the
centres of groups and clusters.

The division of galaxy clusters into samples of cool core (CC) and non
cool core (NCC) systems is well-established \citep[e.g.][]{peres98}.
Recent work has shown both CC and NCC galaxy clusters to show similar,
steep abundance gradients \citep{sanderson08,sivanandam08}. However,
earlier observational work by \citet{degrandi01} indicated that NCC
clusters have flat abundance profiles, in comparison to the steep
abundance gradients seen in CC clusters. The prevalence of merging
systems in the NCC sample of this work led the authors to interpret
mergers as a mechanism for re-distributing metals.

The division into CC and NCC classes has only recently been applied to
study the properties of a sizeable sample of systems in the group
regime \citep{johnson09b}. The origin of NCC clusters remains an open
question, given their short central cooling times \citep{sanderson06},
indicating that either the formation of cool cores in these systems
has been suppressed, for example through pre-heating
\citep[e.g.][]{mccarthy08} or thermal conduction
\citep[e.g.][]{voigt04}, or the cores in NCC systems have been
disrupted by mergers \citep[e.g.][]{allen01b} or AGN heating
\citep[see the review by][]{mcnamara07}. These physical processes
scale differently with system mass, so we can gain significant insight
into the origin of NCC systems by studying NCC groups. The abundance
behaviour in NCC groups has not previously been studied in a sample of
any size, and could prove a useful diagnostic in establishing the
dominant physical processes influencing the ICM.

The layout of the paper is as follows. In Section~\ref{sec:sample} we
describe the group sample and data analysis, in Section~\ref{sec:kT}
we present the mean temperature profile of the CC and NCC groups and
in Section~\ref{sec:abund} we present the abundance profiles of the CC
and NCC groups. We discuss our results in Section~\ref{sec:discuss}
and present our conclusions in Section~\ref{sec:conclude}. Solar
abundances are quoted as those of \citet{and89}.

\section{Group Sample and Spectral Analysis}
\label{sec:sample}
Our sample of 28 galaxy groups is a combination of the groups with the
highest quality \XMM\ data from the Two-Dimensional \XMM\ Group Survey
(2dXGS) sample of \citet{finoguenov06,finoguenov07} and the group
sample of \citet{mahdavi05}. Here we present a summary of the sample
and the data analysis procedures, although we refer the reader to
\citet{johnson09b} for a detailed discussion of the group
sample. Twenty-seven of the groups are situated within z $<$ 0.024,
with the final group at a redshift of 0.037 \citep[RGH\,80,
see][]{mahdavi05}. The data reduction is described in detail by
\citet{mahdavi05} and \citet{finoguenov06,finoguenov07}, but we
present a brief summary of the approach here.
Instead of a traditional annular spectral analysis, spectra were
extracted from regions of contiguous surface brightness and
temperature. This deprojection method
involves no \textit{a priori} assumption of spherical symmetry, but on
the other hand it does not correct for emission from overlying layers
of material. We refer readers to \citet{mahdavi05},
\citet{finoguenov06} and \citet{finoguenov07} for full details.

Results from two spectral analyses are presented here. In the first, 
spectra
were extracted in the range 0.5--3\,\keV\ and fitted with single--temperature
(1T) hot plasma (\textsc{apec}) models to yield
the temperature $T$ and abundance $Z$ in each region. Deprojected values 
of entropy $S$ and pressure $P$ were calculated by assuming a length along the
line-of-sight for each spectral region derived from its
distance to the centre of the system. These results have already been used
to examine the feedback properties of groups \citep{johnson09b}. 

Assuming a single temperature model in regions
where a spread of temperatures is present, as may be the case within cool
cores or where regions of different temperature are projected on top of
one another, can lead to systematic underestimation of the metallicity 
\citep{buote98}. To check for this, a second spectral analysis is
performed, in which spectra are fitted with a 
two--temperature (2T) model, in which abundance is taken to be the same
in both phases. To
better constrain these 2T models, spectra were extracted
across a wider energy range (0.5--7\,\keV). The original motivation
for using a smaller energy range for 1T fits was to
constrain the temperature primarily using the position of the lines, aiming
to reduce bias in the metallicity arising from any distortion in
the continuum. 
We concentrate here on the properties of the abundance profiles of
groups, having already explored the diversity in group properties
such as entropy, and the implications
for feedback processes, in \citet{johnson09b}.

We are entering an era where large galaxy samples can be used to study
the gas properties of galaxy groups \citep[e.g.][]{rasmussen07,sun08}.
In particular, \citet[][hereafter referred to as RP07]{rasmussen07}
derived detailed temperature and abundance profiles for 15 systems, 14
of which were shown to host a CC. The size and diversity of our sample
allows its division into CC and NCC systems \citep[e.g.][]{peres98},
based on the properties of their observed temperature profiles. The
groups in our sample were classified as CCs if the ratio of the
temperature in the radial range 0.1--0.3\,\RF\ to the temperature in
the radial range 0.0--0.05\,\RF\ was found to be greater than 1. This
was found to be a successful discriminator between the systems showing
central temperature drops and those which do not. Although not a
statistically selected sample, the 18~CC and 10~NCC groups allow us to
identify trends in properties based on this segregation -- a first in
the study of galaxy groups.

To negate the effects of the differing sizes of the systems under
consideration, we have scaled radial measurements by \RF\ (measured in
kpc), the radius within which the mean density is equal to 500 times
the critical value. This was defined for the 2dXGS groups by
\citet{finoguenov06,finoguenov07} in the following way,
\begin{equation}
\RF\ = 0.391~\bar{T}^{0.63}h_{70}^{-1},
\end{equation} 
where $\bar{T}$ is the temperature (in keV) measured in the radial
range 0.1--0.3\,\RF\ using a single--tempeature spectral model. The groups 
included in our work from the
\citet{mahdavi05} sample have been re-analysed to extract $\bar{T}$
and \RF.

\subsection{Background Fitting Methodology}

Our approach to the complex issue of removing the \XMM\ background
uses a fitting methodology that allows for the fact that the
background is changing in time and space. The details of the
background treatment adopted in obtaining the \XMM\ results used in
this work are presented by \citet{finoguenov07}. We perform the
standard subtraction of the quiescent background and allow for a soft
component ($T \sim 0.2$~keV) to account for variations in the Galactic
foreground. In order to fully describe the non--X-ray background, up
to two power-laws were fitted in addition to the thermal model. These
power-laws were not convolved with the effective area of the
telescope, achieved using the `/background' model in
\textsc{xspec}. In the X-ray faintest regions of groups this
background component dominates at energies $E\ga 3$~keV. In
combination with the distinct continuum shape of the $\sim 1$~keV
thermal group emission, this allows us to characterise this component
in a robust and unbiased manner. 

When computing uncertainties on fitted source parameters, all other
parameters, including those associated with the fitted background
model, were allowed to reoptimise. Hence, uncertainties on derived $T$
and $Z$ due to the background level are included in our error
budgets. This is a more flexible treatment of the background compared
to that often employed for nearby galaxy clusters, where the
background is inferred using different observations. In fitting
cluster emission, a recent practice has been to add a systematic error
due to the background subtraction. We have refrained from doing so,
since we directly fit for local background shape in the spectral
analysis. However, the faintest groups studied here (and those having
short \XMM\ exposures) cannot be traced to the edge of the \XMM\
field-of-view, and therefore it is not possible to accurately fit the
source spectrum and power-law component in these outer regions. Such
zones are therefore excluded from our analysis.

\begin{figure}
\begin{center}
\includegraphics[width=7cm]{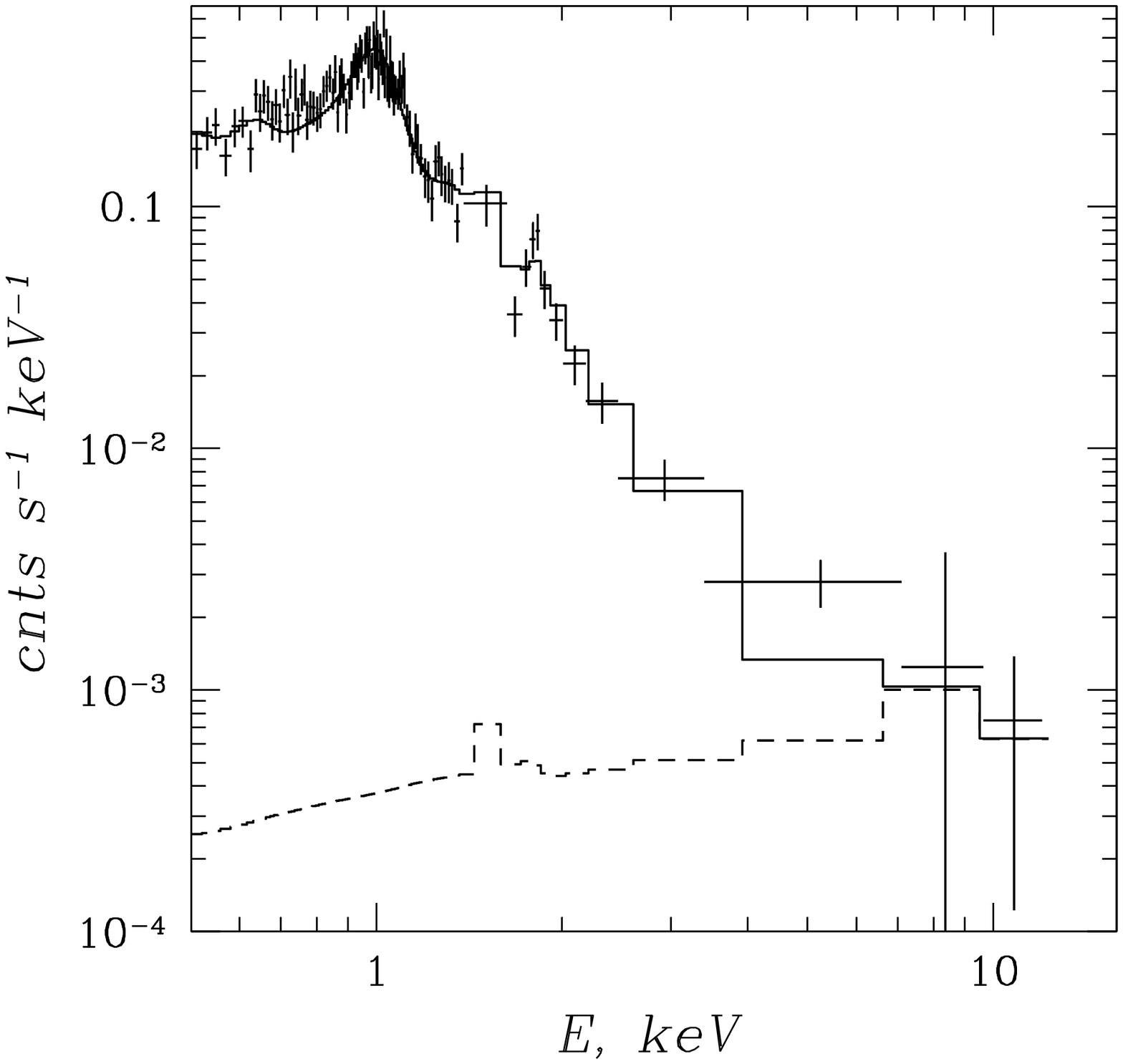}\vspace{-10mm}
\includegraphics[width=7cm]{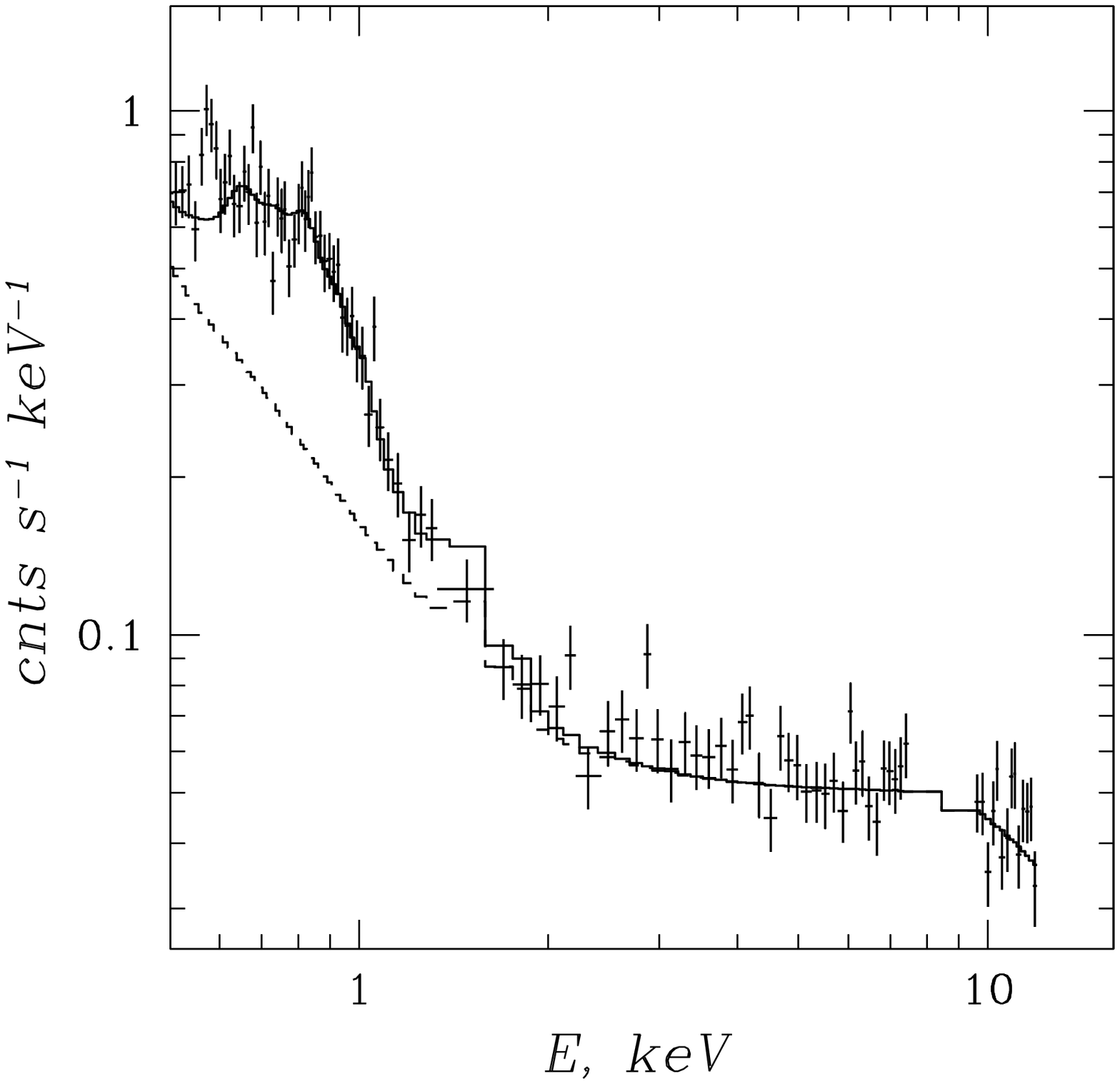}\vspace{-4mm}
  \caption{X-ray spectra (data points) and fitted source + background
  models (solid line) for the groups NGC\,507 (top) and NGC\,5171
  (bottom). The dashed line shows the remaining background component
  once the standard background subtraction has been applied using
  blank sky fields. For both groups, these spectra were extracted from
  the outermost regions for which abundance determinations were
  possible.}
  \label{fig:bg}
\end{center}
\end{figure}

To show the variation in data quality in the outermost regions used
for abundance determinations, we plot in Figure~\ref{fig:bg} examples
of the fitted spectra and background for two of the groups in our
sample (NGC\,507 and NGC\,5171). These were considered to show typical
`best' and `worst' cases, in terms of constraining both the line and
continuum emission in the spectra. The Figure demonstrates that even
in the worst case (NGC\,5171), the line and continuum emission are
well-constrained, and show that the source emission is clearly
discernible from the background, indicating that we are not
over-interpreting the data at large radius. To further test the
robustness of our spectral results and guard against the possibility
that the $\chi^2$--minimisation of our fits would become trapped in
local rather than global minima, we also inspected the $\chi^2$
contour maps in the $T$--$Z$ plane for a number of groups and spectral
regions. In all cases considered, only one miminum was seen, even for
confidence ranges larger than $3\sigma$.

\section{Temperature Profiles}\label{sec:kT}

To illustrate the key differences between the temperature profiles of
CC and NCC groups, we have stacked and scaled radially to \RF\ the
temperature profiles derived from the 1T spectral fits
for each sub-sample to give typical profiles. We
have removed the dependence on the size of the system by dividing the
temperature profiles by the characteristic mean temperature (derived
within the radial range 0.1--0.3\,\RF). We performed a local 
regression `loess' fit to the CC temperature profiles, weighting these fits 
by the inverse variance of the temperature measurement at each point. The 
algorithm fits a two-degree polynomial function using weighted least 
squares in the in the local neighbourhood of each data point. The size of 
the neighbourhood is defined to include a specified proportion of the data, 
which then dictates the smoothness of the resulting fit. The distance to 
each neighbour is used to weight the fit at each point. For more information, 
we refer the reader to \citet{cleveland79} and \citet{cleveland92}.

\begin{figure}
\includegraphics[width=8cm]{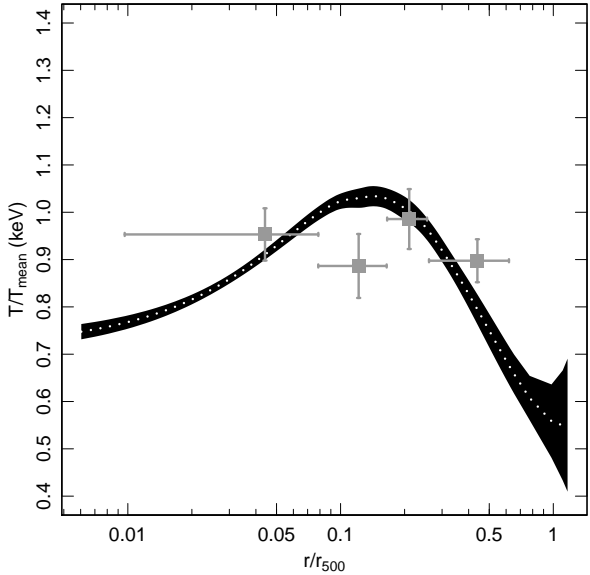}
\caption{The typical temperature profiles from 1T spectral 
  fits(scaled by the mean group
  temperature, measured in the radial range 0.1--0.3\,\RF) for CC
  groups (black confidence region and white dotted line) and for NCC
  groups (filled squares with error bars). The typical temperature
  profile in the CC case is derived via a weighted local regression
  fit, where the associated confidence region shows the standard 
  error on the fit. The number of groups contributing to the
  CC profile falls to two at a radius of 0.6\,\RF, so the
  interpretation of the profile beyond this radius should be treated
  with caution. Due to poorer statistics in the NCC case, we have
  stacked the profiles into four radial bins, calculating the mean
  scaled temperature (grey squares) and the standard error on the mean scaled
  temperature (grey error bars).}
\label{fig:stack}
\end{figure}

However, due to the diversity in the
temperature profile properties of the NCC groups, the regression fit
in this case was unsuccessful, and instead we divided the data into
four radial bins, each containing between 24 and 25 data points. To 
make a direct comparison with the CC profile, we calculated a weighted 
mean of the scaled temperature points in each radial bin, and also calculated
the standard error on the mean scaled temperature (i.e. the rms scatter of the $n$ values falling in each bin,
divided by $\sqrt{n}$), to show the typical 
behaviour of the NCC temperature profiles. These temperature profiles for 
the CC and NCC
groups are shown in Figure~\ref{fig:stack}. 
The standard error on the regression fit to the CC groups inflates at both
small and large radius; this is simply due to the lower number of data points
in these regions. The larger standard error at radii greater than 0.5\,\RF\ reflects the increased variance arising from larger measurement errors in this
region of lower surface brightness. Figure~\ref{fig:stack} indicates a higher 
degree of consistency between the temperature profiles of CC systems, shown via
the narrow standard error in comparison to the NCC systems.

\section{Abundance Profiles}\label{sec:abund}
\begin{figure*}
\includegraphics[width=17cm]{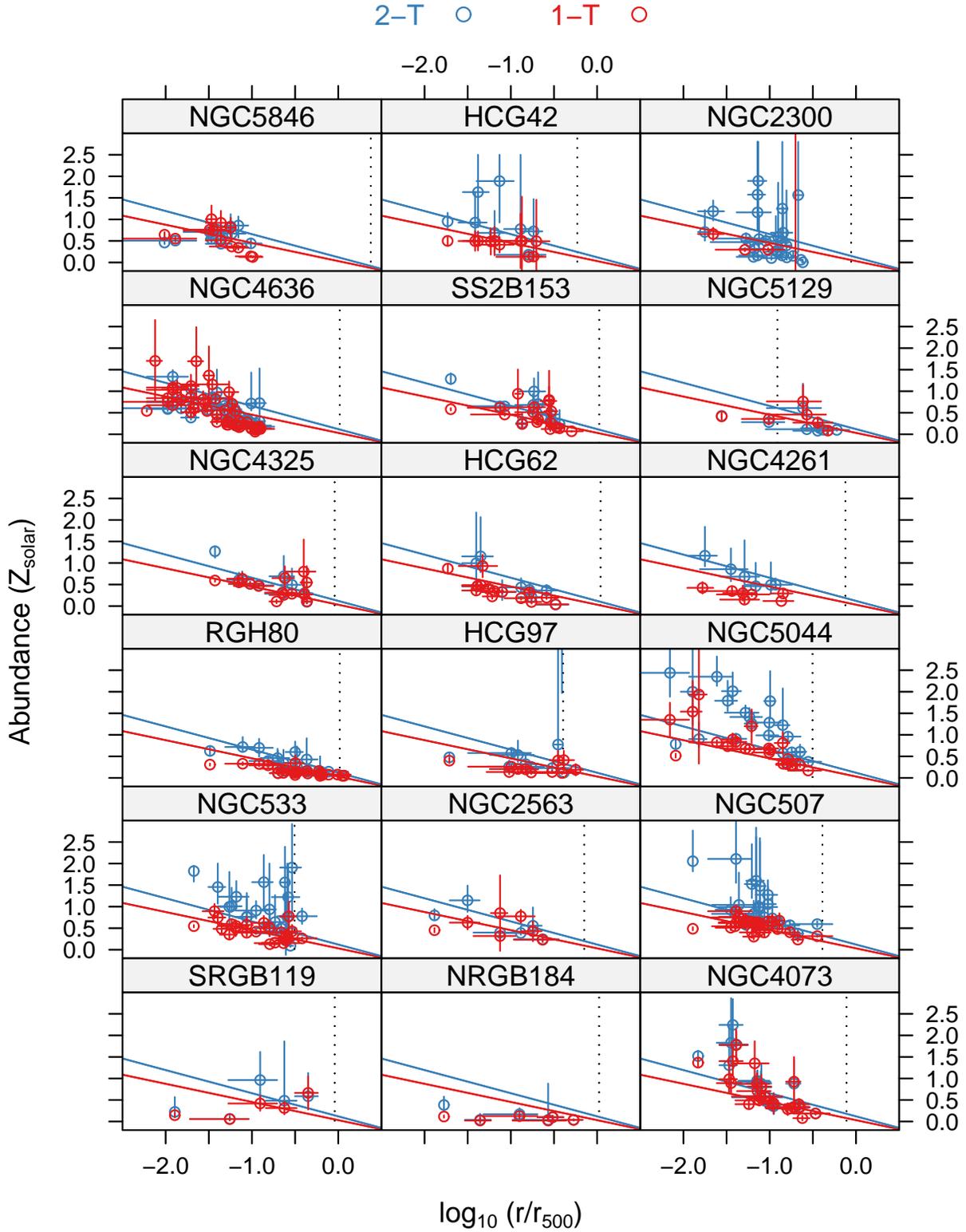}
\caption{The abundance profiles of the CC groups in the 2dXGS
  group sample, shown in log--log space. The colours denote 
  1T (red) and 2T (blue) spectral fits.
  Vertical error bars are measurement
  errors and horizontal error bars show the width of each radial
  bin. The vertical dotted lines show the \XMM\ field-of-view of
  16$^{\prime}$ for all groups except NGC\,4636 and NGC\,5044, where
  the offset of the pointings shifts the outer boundaries to
  18$^{\prime}$ and 17.7$^{\prime}$ respectively. The solid lines
  show the results of linear model fits to all CC groups, for
  both the 1T (red) and 2T (blue) 
  spectral fits.}
\label{fig:Z-CC}
\end{figure*}

\begin{figure*}
\includegraphics[width=17cm]{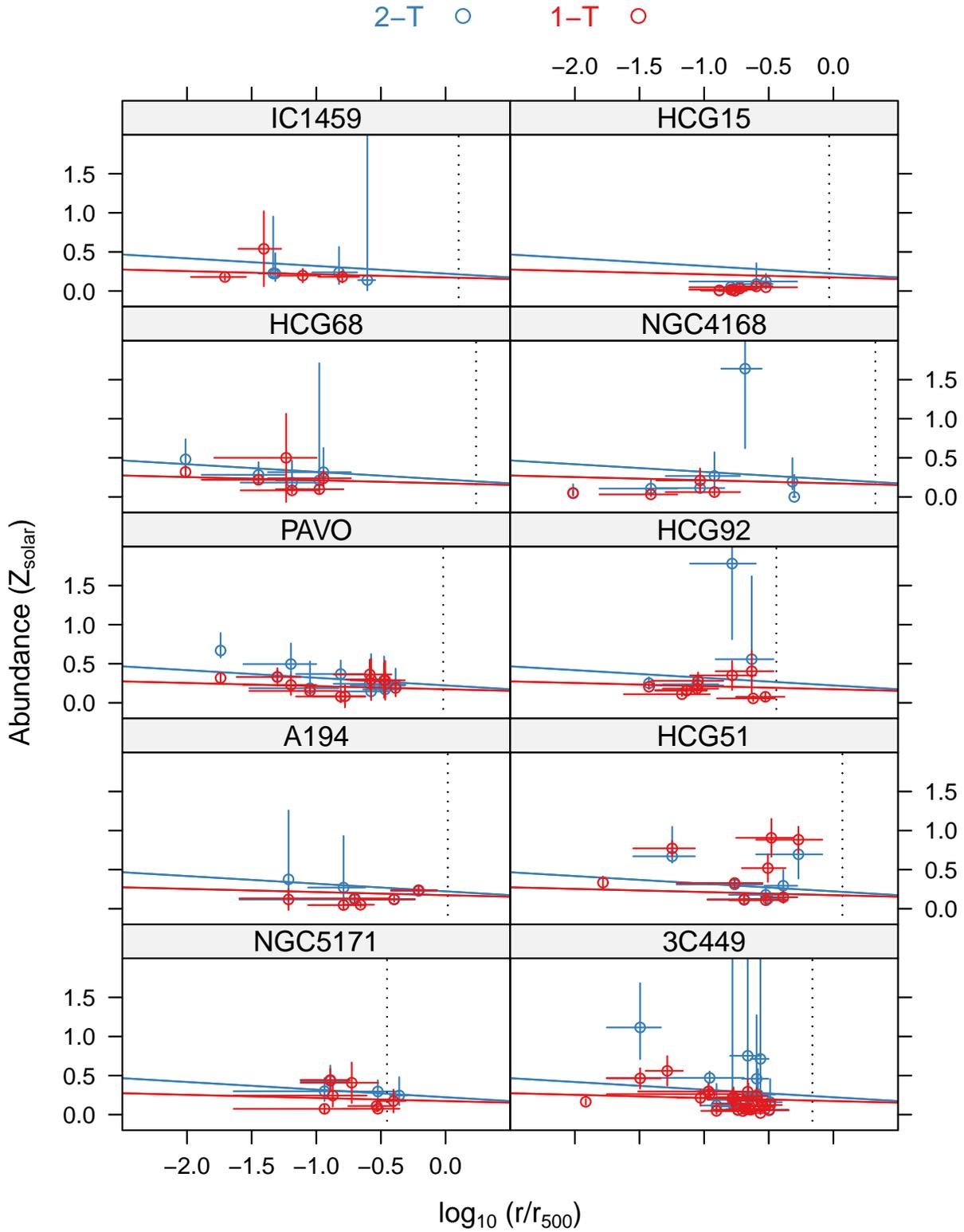}
\caption{The abundance profiles of the NCC groups in the 2dXGS
  group sample, shown in log--log space. The colours denote 
  1T (red) and 2T (blue) spectral fits.
  Vertical error bars are measurement
  errors and horizontal error bars show the width of each radial
  bin. The vertical dotted lines show the \XMM\ field-of-view of
  16$^{\prime}$ for all groups. The solid lines
  show the results of linear model fits to all NCC groups, for
  both the 1T (red) and 2T (blue) 
  spectral fits.}
\label{fig:Z-NCC}
\end{figure*}

\citet{finoguenov99} and \citet{buote00} showed CC galaxy groups to
have a central iron peak, a result confirmed for a larger sample of
groups by RP07. However, the abundance profiles of NCC groups have not
been previously investigated. Figures \ref{fig:Z-CC} and 
\ref{fig:Z-NCC} show the abundance profiles from both 
1T and 2T spectral fits, for both the CC
and NCC groups in the 2dXGS sample. In both figures, the horizontal error 
bars show the radial width of each bin, and the vertical error bars show the
measurement error derived from \textsc{xspec}, which takes into
account the uncertainty on the modelled background. 

As expected, the 2T models give higher metallicity, especially
in regions from the which the spectrum is not well-represented by an 
isothermal plasma. This is especially the case within cool cores, where
the steep gradient results in regions of differing temperature being
projected on top of one another, and also thermal instability may result
in multiphase gas \citep{buote98}. We note that the results from 
1T and 2T models converge at large radii,
and the abundance offset between the two models is more significant in
the case of CC groups, where the inferred abundance gradient is
steepened by the use of a 2T model. NGC\,5044 (see 
Figure~\ref{fig:Z-CC}) provides an especially striking example.

\begin{figure}
\includegraphics[width=8cm]{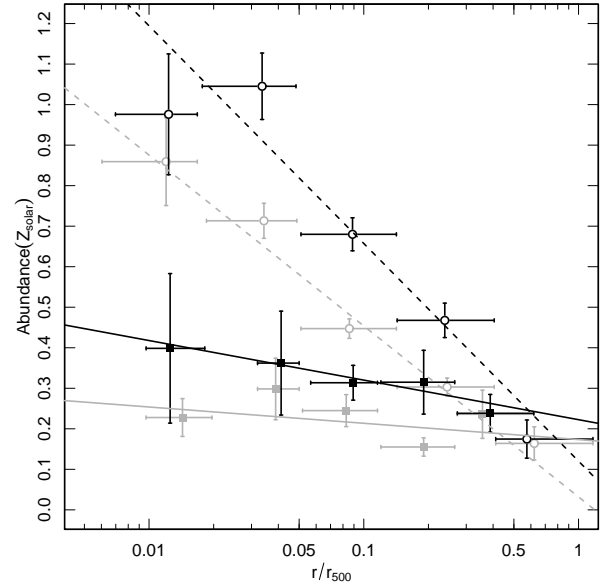}
\caption{The stacked abundance profiles of the groups, divided into
  the CC (open circles) and NCC (filled squares) sub-samples. The results
  of the 1T fits are shown in grey, and the results of
  the 2T fits are shown in black. The
  vertical error bars are the standard error on the mean profile,
  and the
  horizontal error bars show the radial width of each bin. The dashed
  and solid lines are linear fits to the whole data set of abundances
  from the CC and NCC groups respectively, colour coded for the 
  1T (grey) and 2T (black) fits. }
\label{fig:Z}
\end{figure}

Abundance profiles in CC and NCC groups can be compared by 
stacking results from the groups in each subsample. We divided the 
radial range of the 
data for the CC and NCC systems into five equally spaced bins (in log space), 
and within each radial bin determined the mean abundance from the ensemble
of results falling within this bin, and a standard error from their scatter.
The result is
shown in Figure~\ref{fig:Z}. Results were derived separately for 
1T and 2T fits, to illustrate the impact
of the 2nd temperature component on the fitted abundances. The horizontal 
error bars in Figure~\ref{fig:Z} show the width of the radial bins used in the
stacking analysis. The vertical error bars show the standard error
for each radial bin. This is larger for 2T models, since
the larger number of free parameters in the model leads to greater
scatter in the results.

Figure~\ref{fig:Z} also shows the results of performing a linear
regression on the original unbinned data for the CC and NCC samples
separately. No statistical weighting has been applied when calculating 
these regression lines, since 
the variance about the mean profile is dominated by system-to-system 
variations, rather than statistical scatter. 
We fit linear models in log--linear
space using the \rr\ function \textsc{`lm'} for linear regression,
finding the following relations, using 1T models, for the CC groups,
\begin{equation}
\label{eqn:CC-fit}
Z_{\rm CC} =
-0.42~(\pm0.04)~\mathrm{log}\left(\frac{r}{\RF}\right)+0.03~(\pm0.04)
\mbox{ Z}_\odot ,
\end{equation}
and for the NCC groups,
\begin{equation}
Z_{\rm NCC} =
-0.04~(\pm0.05)~\mathrm{log}\left(\frac{r}{\RF}\right)+0.17~(\pm0.05)
\mbox{ Z}_\odot .
\end{equation}
Under the 2T spectral models, the fitted trend for the CC
groups is,
\begin{equation}
\label{eqn:CC-fit2}
Z_{\rm CC} =
-0.54~(\pm0.07)~\mathrm{log}\left(\frac{r}{\RF}\right)+0.12~(\pm0.07)
\mbox{ Z}_\odot ,
\end{equation}
and for the NCC groups,
\begin{equation}
Z_{\rm NCC} =
-0.10~(\pm0.10)~\mathrm{log}\left(\frac{r}{\RF}\right)+0.22~(\pm0.09)
\mbox{ Z}_\odot .
\end{equation}

Clearly, the CC systems exhibit a much steeper abundance gradient, 
compared to the NCC systems. The fit to the NCC profiles is consistent with
being flat within the quoted standard errors. This is true for both the
fits from the 1T models and the 2T models. 
In the case of the CC profiles, results from 2T fits give
a steeper slope than 1T models.

The group HCG\,51 has a relatively high abundance in the 1T 
fits
($\sim$0.5--0.8~Z$_\odot$) at radii between 0.3\,\RF\ and 0.5\,\RF,
which affects the calculation of the mean abundance in the outermost
bin of the NCC profile in this case. To assess the impact of HCG\,51 on 
this mean profile, we re-calculated the mean abundance in the outer bin,
excluding HCG~51. This reduced the mean value to 0.14$\pm$0.02$\Zsol$ from 
0.24$\pm$0.06$\Zsol$. Other
bins are only marginally affected by the exclusion of HCG\,51. Fitting
a straight line model in log--linear space to the NCC groups, excluding all
HCG\,51 data, we find
\begin{equation}
Z_{\rm NCC} =
-0.09~(\pm0.04)~\mathrm{log}\left(\frac{r}{\RF}\right)+0.10~(\pm0.04)
\mbox{ Z}_\odot ,
\end{equation}
yielding a slope that is non-zero (within a 95\% confidence interval),
but still much shallower than that found in the CC systems. 
In the case of 2T spectral fits, excluding 
HCG\,51 gives a mean profile for NCC groups
\begin{equation}
Z_{\rm NCC} =
-0.10~(\pm0.11)~\mathrm{log}\left(\frac{r}{\RF}\right)+0.21~(\pm0.10)
\mbox{ Z}_\odot ,
\end{equation}
with a slope which is consistent with zero within the 1$\sigma$ error.

The presence of a central abundance peak is consistent with being
built from the products of type Ia supernovae occurring in the central
galaxy, in both groups \citep{rasmussen09} and clusters
\citep{david08}. However, the typical optical extent of the central
group galaxy is only 0.05\,\RF, indicating that the metallicity
profile of the CC and NCC systems continues to fall well outside the
central galaxy. This suggests that metals have been expelled from the
central galaxy into the surrounding intracluster medium. We estimate a
mean gas mass weighted metal fraction for the CC and NCC groups by
summing the product of the metallicity and gas mass over a series of
radial shells, and dividing by the total gas mass contained within
these shells. The gas mass was derived from $\beta$--model fits to the
gas density profiles \citep[see][for more information]{johnson09b}.
The calculation of the metal fraction was limited to within 0.3\,\RF\ to
ensure consistent radial coverage between the groups, and this is also
the radius where the CC and NCC profiles begin to converge within the
uncertainties in Figure~\ref{fig:Z}.

Applying this analysis to the results from the 1T models,
we find a mean metal fraction for the CC groups of $0.29\pm 0.03$,
where the error quoted is the standard error on the mean metal 
fraction, whilst for the NCC groups the mean metal fraction is 
$0.16\pm 0.02$, almost a factor of two lower. 
In the case of the 2T models, abundances are higher, but the
mean metallicity in CC systems within 0.3\,\RF\ is still 
double that for NCC groups.
A key question is whether the central metals seen in CC
groups are missing in NCC groups, or whether they have just been mixed
out to larger radii. For most of the CC groups, we do not have good
metallicity estimates outside 0.5\,\RF. We therefore adopt an
abundance of 0.18 solar outside 0.3\,\RF, corresponding to the average
behaviour. We then find the mean total metal mass inside 0.3\,\RF\ to
be typically one half of that in the radial range 0.3\,\RF--\RF.  This
is not an insignificant fraction, so mixing out the central metal peak
should have a substantial impact on the outer regions.  Thus, if the
central metal peak has been mixed out to a large radius in NCCs, it
would be expected to lead to a significant rise in metallicity in the
outer regions, compared with what is seen in CC groups.  Within the
limits of our data, there is no evidence for this, except in the case
of HCG\,51. However, better quality spectral data extending to large
radii is required to firmly establish whether or not a mixing scenario
is viable.


\subsection{CC/NCC definition}
The CC/NCC definition employed by \citet{johnson09b} compared the
temperature profile behaviour in the radial range 0--0.05\,\RF\ with
that in the radial range 0.1--0.3\,\RF. Therefore, when we classify a
group as a CC, we are referring to a classic cool core system,
equivalent to the LCC systems of \citet{sun08}, with a core typically
extending to $\sim$0.1\,\RF.  With this definition, any groups showing
a central temperature drop on very small radial scales ($<$ 10\,\kpc),
would be classified as NCC. Such behaviour is seen in $\approx
20$~per~cent of systems in the \Chandra\ sample of \citet{sun08},
termed `coronae' class systems by Sun et~al. They possess a small cool
region lying within the central galaxy. None of the CC systems in the
\Chandra\ sample of RP07 shows temperature drops on such small radial
scales. Most importantly for this work is the observation by
\citet{sun08} of compact cool regions within 10\,\kpc\ in 3C~449 and
HCG\,51, classified in this work as NCC systems. We tested the
susceptibility of our results to the applied CC/NCC definition in
these two systems by changing their designation and determining the
abundance gradients of the resulting stacked CC and NCC profiles.
Again fitting a linear model in log--linear space to the results from
both the 2T models and the 1T models, we
find the slope and intercept of the abundance profiles to be consistent
with the original profiles within the stated errors, for both the CCs and
the NCCs. The fitted intercepts and slopes are shown in Table \ref{table:fits}.
Therefore, even if 3C~449 and
HCG\,51 have their CC status reclassified, the results from the
stacking analysis are not significantly affected.

\begin{table}
  \centering
  \caption{The fitted intercepts, slopes and errors for models fitted to the
  1T and 2T results with 3C449 and HCG~51
  reclassified as CC systems.}
  \label{table:fits}
  \begin{tabular}{cccc}
  \hline
      &  & NCC & CC  \\
  \hline
  1-T & Slope     & -0.05$\pm$0.05 & -0.06$\pm$0.12\\
      & Intercept & 0.13$\pm$0.05  & 0.03$\pm$0.03\\
  2-T & Slope     & -0.06$\pm$0.12 & -0.56$\pm$0.06\\
      & Intercept & 0.24$\pm$0.12  & 0.09$\pm$0.07\\
   \hline
  \end{tabular}
\end{table}
%

\subsection{Comparison to \citet{rasmussen07}}
Although the detailed abundance profile behaviour of NCC groups has
not been previously examined, we can compare the behaviour of the CC
abundance profiles here with those of RP07. The latter work benefits
from the higher spatial resolution available with \Chandra\ data, so
we cannot draw inferences here on the presence of the central (within
0.01\,\RF) drop in abundance seen by RP07. We can however, compare the
overall trend seen in the CC systems. There is an overlap of 10 CC
systems between this work and the sample of RP07, which further allows
a comparison of the spectral analysis methods for those systems in
common. RP07 fitted their spectra with 2T models
wherever these gave a significant improvement in fit, which was often
the case within the cool core. We therefore show the comparison with
both our 2T and 1T results.

\begin{figure}
\includegraphics[width=8cm]{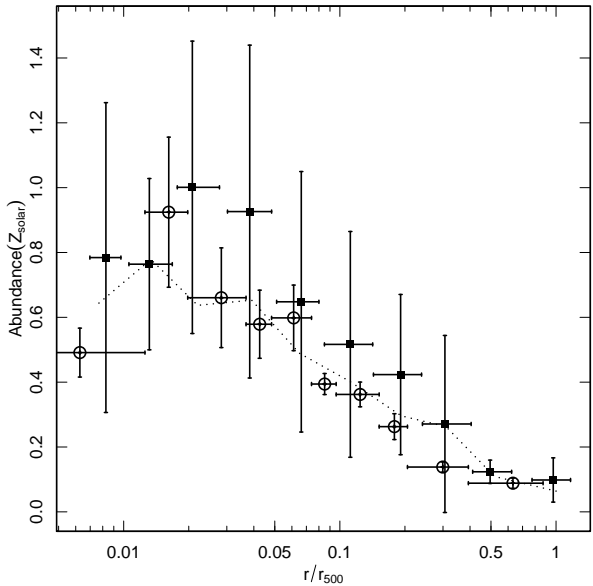}
\caption{The stacked abundance profiles of the CC groups from our
  sample using the 2T analysis (solid squares), derived 
  from the median abundance in each
  radial bin, with error bars that correspond to the median absolute
  deviation to show the degree of scatter in each bin. Open circles
  show the results of \citet{rasmussen07}, where we have re-scaled
  their abundances by 1.48 to match the \citet{and89} abundances
  presented here, and have re-scaled the \RF\ values of
  \citet{rasmussen07} to allow for differences in the method of
  calculation. The dotted line shows the median abundances from the
  1T spectral fits.}
\label{fig:ras}
\end{figure}

To enable a fair comparison, we convert the \cite{grevesse98}
abundances presented by RP07 to \citet{and89}; this requires dividing
the former by a factor of 1.48 (as described by RP07). A further
correction is required to allow for the difference in the method of
calculating \RF\ in the two samples, as the RP07 \RF\ values are
typically $\sim$1.14 times greater than the values used here. We have
scaled the \RF\ values of the RP07 sample down by this factor to
compare to our work. We have again stacked the abundance profiles for
the CC groups, this time increasing the number of radial bins to allow
a more thorough comparison with the profile of RP07. We also now
calculate the median in each bin, to avoid any bias from an individual
group influencing the overall profile. Although the mean trends are
well-established (see Figure~\ref{fig:Z}), individual groups do show
some deviations from these mean profiles. To indicate the degree of
scatter in each bin in Figure~\ref{fig:ras} we plot (as error bars)
the median absolute deviation in each radial bin.

Figure~\ref{fig:ras} shows the stacked abundance profile for the CC
groups from the current sample under the 2T analysis, shown 
as solid squares, and the stacked
abundance profile from the sample of RP07, shown as open circles. The
points from this analysis tend to be higher than those of RP07, however
the spread of abundances in each bin is much larger. It is worth noting that 
RP07 use \textsc{vapec} spectral
models in \textsc{xspec}, so that the abundance shown in
Figure~\ref{fig:ras} is actually the iron abundance. Within the RP07
sample, the difference 
between iron abundance and the mean metallicity is at most 15\%. Since
RP07 used a mixture of 1T and 2T spectral fits within their group cores, 
we also include our 1T profile in the Figure. In general, the shape of 
our mean abundance profile is consistent with that of RP07, though the 
match is closer for our 1T fits.

\subsection{Comparison to clusters}

Recent work on the abundance profiles of CC and NCC clusters has shown
them to exhibit very similar profiles. For example, an analysis of
\Chandra\ data by \citet{sanderson08} showed NCC clusters to show a
similar decline with radius to CC clusters, and \citet{sivanandam08}
find steep abundance gradients in 9/12 clusters, of which only 4 are
CCs. At intermediate redshift (0.1 $<$ z $<$ 0.3), \citet{baldi07}
showed the abundance profiles of CC clusters to rise above those of
NCC systems within 0.1\,$r_{180}$, however, outside this radius, they
found no significant difference in the profiles of CC and NCC
clusters. Earlier work by \citet{degrandi01} and \citet{degrandi04}
with \BSAX\ data showed almost flat abundance profiles in NCC
clusters, compared to steep abundance profiles in CC
clusters. However, the NCC sample used in the \BSAX\ study consisted
of well known merging clusters, which probably accounts for the
different profiles in these compared to more recent studies of NCC
clusters (S. Molendi, private communication).

\begin{figure}
\includegraphics[width=8cm]{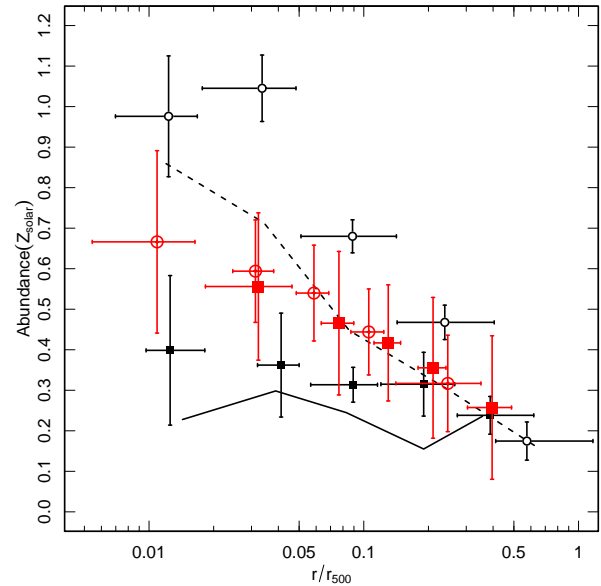}
\caption{The stacked abundance profiles of the CC groups (open
  circles) and NCC groups (filled squares) from the 2T
  spectral fits, with
  vertical error bars showing the standard error on the mean abundance, and
  horizontal error bars showing the radial width of the bin. For
  comparison, we show the mean abundance profiles from the cluster
  sample of \citet{sanderson08} in red, with open circles denoting CC
  clusters and filled squares denoting NCC clusters. The horizontal
  error bars show the radial width of each bin, and vertical error
  bars show the standard deviation in each bin. We have re-scaled the
  abundances of \citet{sanderson08} to match the \citet{and89}
  abundances presented here. The dashed and solid lines show the 
  mean results of
  our 1T spectral fits for CC and NCC groups respectively.}
\label{fig:clusters}
\end{figure}

In Figure~\ref{fig:clusters} we compare the mean abundance profiles
for the CC and NCC groups in our sample with the mean abundance
profiles of a sample of 20 clusters, presented by
\citet{sanderson08}. The latter profiles have also been split into CC
and NCC categories, and result from single--temperature fits to spectra
extracted from annuli. Figure~\ref{fig:clusters} shows the striking
similarity between the abundance profiles of CC and NCC clusters, in contrast
to the situation in groups, where
CC and NCC systems have distinctly different profiles. Comparing CC groups
to CC clusters shows CC groups to have a higher central peak, whether 1T
or 2T models are employed,
which may be explained by the increased dominance of the brightest
group galaxy (BGG) in lower mass systems \citep[e.g.][]{lin04b}. NCC groups
however have considerably flatter abundance profiles than 
NCC clusters. We will return to 
this point in Section~\ref{sec:discuss}.

%

\subsection{Low abundance systems}\label{sec:ind-Z}
Motivated by the challenge of determining the main driving factors
that lead to the apparent bimodality in the abundance profiles
presented in Section~\ref{sec:abund}, we have looked for systems that
buck these mean trends. We find four systems to have very low
abundances of less than 0.2~Z$_\odot$ across the measured radial
range. Three of these groups are NCCs (HCG\,15, NGC\,4168 and A\,194)
and one is a CC system (NRGb\,184). Given the role of the BGG in
establishing the central peak in abundance
\citep[e.g.][]{david08,rasmussen09}, we hypothesise that these low
abundance systems have relatively small (i.e.\ lower stellar mass)
brightest group galaxies, such that the relative injection of metals
from the BGG is low.

We can assess this by calculating the ratio of the $K$-band luminosity
of the BGG to the total gas mass within 0.3\,\RF.  We estimate the
$K$-band luminosity of the BGGs using $K_{20}$ magnitudes from 2MASS
\citep{skrutskie06}, assuming a $K$-band absolute magnitude for the
Sun of 3.39 \citep{kochanek01}. The only system for which we do not
have a $K$-band magnitude for the BGG is HCG\,51. The gas mass comes
from $\beta$-model fits to the gas density profiles of the individual
groups \citep[see][for more information]{johnson09b}. Simply measuring
the mean \ensuremath{L_{K}({\rm BGG})/M_{\rm gas}} of these low
abundance systems we find $L_K / M_{\rm gas} = 0.38 \pm
0.07$~L$_{\odot,K}$~M$_{\odot}^{-1}$, compared to the remainder of the
sample for which $L_K / M_{\rm gas} = 0.74 \pm
0.12$~L$_{\odot,K}$~M$_{\odot}^{-1}$, where quoted errors are the
standard error on the mean.

This indicates that the low abundance systems do indeed have a smaller
ratio of stellar to gas mass within 0.3\,\RF.  Is this sufficient to
account for their lower metallicity?  To assess this, we calculate the
ratio of integrated iron to total stellar mass, to see whether it is
abnormally low in the low abundance groups. Computing the product of
gas mass and metallicity summed over a series of radial shells out to
0.3\,\RF, and dividing by the $K$-band luminosity of the BGG, gives an
indication of the metal contribution from the BGG. In the low
abundance systems, the mean of this ratio of `metal mass' to
$L_{K,{\rm BGG}}$ is $0.19\pm 0.03$~Z$_\odot$~M$_\odot$/L$_{\odot,K}$,
whilst for the remainder of the systems, it is $0.59\pm
0.08$~Z$_\odot$~M$_\odot$/L$_{\odot,K}$, where the errors quoted are
the standard errors on the mean. This shows that the lower stellar
mass of the BGGs in the low abundance groups is not sufficient to
account for their low metal mass -- the ratio of metal mass to stellar
mass is actually lower in these systems.  The same conclusion is
arrived at if we allow for the possible contribution of non-central
group galaxies to the metal mass within 0.3\,\RF.  Using the total
$B$-band luminosities within \RF\ from \citet{johnson09b} to normalise
metal mass we obtain a ratio for the low abundance systems of $0.21\pm
0.06$~Z$_\odot$~M$_\odot$/L$_{\odot,B}$, whereas the mean ratio is
$0.91\pm 0.12$~Z$_\odot$~M$_\odot$/L$_{\odot,B}$ for the remainder of
the sample. We conclude that the member galaxies in these low
abundance groups contribute an unusually low metal mass to the ICM
within $0.3r_{500}$.

\subsection{AGN activity}\label{sec:AGN}
A statistical study of the effects of AGN activity on the gas
properties of galaxy groups by \citet{jetha07} showed that although
AGN may have an impact on the \textit{local} gas properties, the large
scale gas structure appears not to be significantly affected. The
sample used by \citet{jetha07} was biased towards hotter systems with
larger X-ray luminosities than ours. The majority of systems in the
sample of \citet{jetha07} show a temperature decline within 0.1\,\RF\
(see figure~4 in \citealt{jetha07}), indicating that their sample is
dominated by CC groups. Here we have the advantage that we can
consider the effects of AGN activity on NCC systems as well.

A study of the effects of feedback on the 2dXGS sample
\citep{johnson09b} concluded that AGN are probably the dominant source
of feedback, rather than supernovae, due to the lack of extra metals
in systems with higher levels of feedback, but that much of this
feedback might have taken place at earlier epochs.  To investigate the
effects of {\it current} AGN activity in these systems, we have
extracted 1.4\,GHz radio fluxes for the BGGs from the references shown
in Table~\ref{table:main}, primarily through the NASA/IPAC
Extragalactic Database (NED). Radio fluxes were not available in all
cases, and we further searched the NRAO VLA Sky Survey
\citep[NVSS;][]{condon98} around the BGG position for radio
sources. The radio fluxes were converted to radio luminosities, and
are shown in Table~\ref{table:main}. The sample with available radio
power estimates consists of 14 CC systems and 6 NCC systems.

\begin{table}
  \centering
  \caption{The mean group temperatures (measured in the region
  0.1--0.3\,\RF) from 1T spectral fits, and the 1.4~GHz 
  radio luminosity of the BGG. The
  final column shows whether the group was classified as CC or NCC by
  \citet{johnson09b}.}
  \label{table:main}
  \begin{tabular}{ccccc}
  \hline
     Group & \ensuremath{\bar{T}} & log\,\LR\ & Radio ref. & CC/NCC\\
           & (keV) & (W~Hz$^{-1}$) & & \\
  \hline
  3C449 & 1.28$\pm$0.02 & 24.31 & C02 & NCC\\
  A\,194 & 1.01$\pm$0.15 & 23.85 & C02 & NCC\\
  HCG\,15 & 0.62$\pm$0.04 & 21.70 & C02 & NCC\\
  HCG\,42 & 0.75$\pm$0.19 & 21.09 & C98 & CC\\
  HCG\,51 & 1.16$\pm$0.13 & -- & -- & NCC\\
  HCG\,62 & 1.06$\pm$0.02 & 21.51 & C05 & CC\\
  HCG\,68 & 0.69$\pm$0.09 & 21.91 & C02 & NCC\\
  HCG\,92 & 0.79$\pm$0.24 & -- & -- & NCC\\
  HCG\,97 & 1.20$\pm$0.05 & -- & -- & CC\\
  IC\,1459 & 0.59$\pm$0.03 & 23.02 & C98 & NCC\\
  NGC\,507 & 1.34$\pm$0.01 & 22.77 & C02 & CC\\
  NGC\,533 & 1.26$\pm$0.01 & 22.30 & C02 & CC\\
  NGC\,2300 & 0.75$\pm$0.01 & 20.47 & C02 & CC\\
  NGC\,2563 & 1.31$\pm$0.05 & -- & -- & CC\\
  NGC\,4073 & 1.87$\pm$0.05 & -- & -- & CC\\
  NGC\,4168 & 0.77$\pm$0.31 & 20.99 & C02 & NCC\\
  NGC\,4261 & 1.11$\pm$0.02 & 24.60 & C02 & CC\\
  NGC\,4325 & 1.01$\pm$0.01 & -- & -- & CC\\
  NGC\,4636 & 0.77$\pm$0.01 & 20.97 & C02 & CC\\
  NGC\,5044 & 1.21$\pm$0.01 & 21.66 & C05 & CC\\
  NGC\,5129 & 0.95$\pm$0.03 & 22.01 & C02 & CC\\
  NGC\,5171 & 1.21$\pm$0.05 & -- & -- & NCC\\
  NGC\,5846 & 0.69$\pm$0.01 & 21.36 & C02 & CC\\
  NRGb\,184 & 1.37$\pm$0.09 & 23.92 & C02 & CC\\
  Pavo & 0.77$\pm$0.12 & -- & -- & NCC\\
  RGH\,80 & 1.16$\pm$0.02 & 23.40 & C98 & CC\\
  SRGb\,119 & 1.34$\pm$0.07 & 23.90 & C02 & CC\\
  SS2b\,153 & 0.83$\pm$0.01 & 21.66 & C02 & CC\\
  \hline
  \end{tabular}
 \begin{list}{}{}
  \item[References:]
  \item[C98] -- \citet{condon98}
  \item[C02] -- \citet{condon02}
  \item[C05] -- \citet{croston05}
  \end{list}
\end{table}

\begin{figure}
\includegraphics[width=8cm]{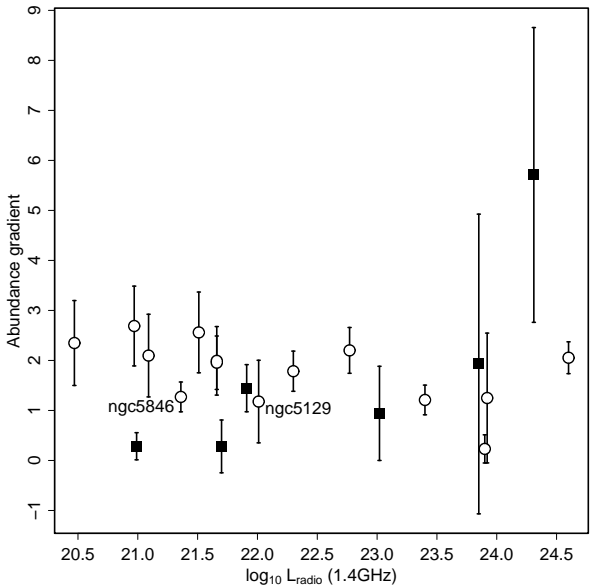}
\caption{The measured abundance gradient within 0.2\,\RF\ from the 2T
  spectral fits (see text) as a function of
  the logarithmic 1.4~GHz radio power of the brightest group galaxy.
  Open circles show CC systems, and filled squares show NCC systems.
  NGC\,5129 and NGC\,5846 are marked due to the difference in methodology for
  the calculation of the abundance gradient.}
\label{fig:Z-radio}
\end{figure}

The observed flatter abundance profiles in NCC galaxy groups compared
to CC groups suggests that a mixing process may be affecting the gas
distribution. If the source of this mixing were AGN, we might expect a
correlation between a flatter abundance gradient and the presence of a
powerful radio source. The 1.4\,GHz radio power measures current AGN
activity rather than recent activity, so the possibility of a time lag
needs to be borne in mind. 
In Figure~\ref{fig:Z-radio} we plot the
observed abundance gradient within the group core
from the 2T fits versus the 1.4\,GHz radio luminosity for
the systems where this latter measurement was available, separating
the systems into CC and NCC groups. We specify the core abundance gradient
as the ratio of the mean abundance measured within 0.05\,\RF\ to that
measured in the 0.1--0.2\,\RF\ radial range. This choice is motivated
by the observation of approximately flat abundance profiles inside
0.05\,\RF\ in CC groups, observed by RP07. This also allows the
abundance gradients to be measured across the same radial range in all
systems. For the group NGC\,5129, no data are
available in the range 0.1--0.2\,\RF, so we take the mean abundance in
the range 0.2--0.3\,\RF\ as the outermost measurement. Given the drop
in abundance with radius in the CC systems, this should lead to an
overestimate of the true abundance gradient by $\sim$35~per~cent. The group
NGC\,5846 also does not contain any data in the larger radial range,
so we use the outermost abundance measurement instead, leading to a 
potential under-estimate of the abundance gradient. These
data points are identified in Figure~\ref{fig:Z-radio}. Two of the NCC systems
with measured radio luminosities have no abundance measurements within
0.05\,\RF. In these cases, we adopt a value within 0.05\,\RF\ equal to
the innermost measurement available. This assumes the innermost
abundance profile is flat, which from Figure~\ref{fig:Z} is a
reasonable assumption. In this scheme, a larger number on the y-axis
in Figure~\ref{fig:Z-radio} indicates a steeper abundance gradient.

The radio luminosities of the brightest group galaxies in the NCC
groups cover a similar range to those in the CC groups. It is therefore
immediately clear that current AGN activity is not responsible for the 
observed difference in abundance distribution between CC and NCC groups. 
For the CC groups, we find anticorrelation at the 95\% confidence level 
between the core abundance gradient and radio power. 
A correlation test yields $\tau =-$0.4, with a p-value of 
0.04. In the case of NCC groups, there is 
weak evidence for a positive correlation, but this is driven by the
two groups with highest radio luminosity, which have very poorly
determined abundance gradients. Better data are therefore required to 
draw any conclusions about any relationship beween radio power
and abundance distribution in NCC systems.

\begin{figure}
\includegraphics[width=8cm]{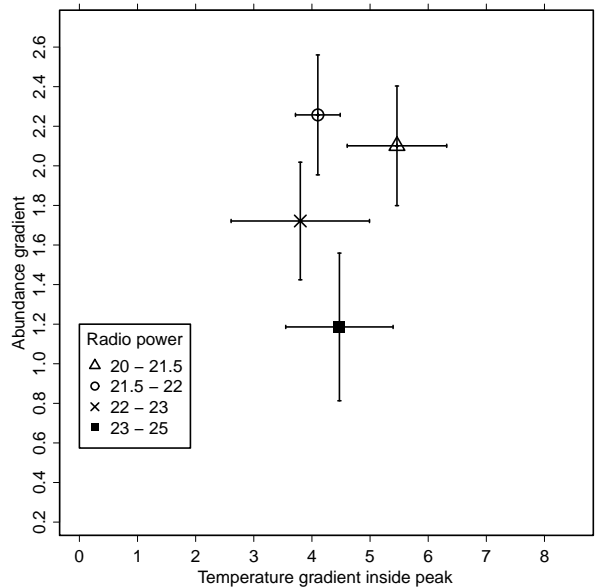}
\caption{The abundance gradient (measured as explained
  in Section~\ref{sec:AGN}) versus the temperature gradient within the
  temperature peak from \citet{johnson09b}, for the CC systems. The 14 
  CC groups with
  available radio powers have been binned by the logarithm of their
  1.4\,GHz radio power, as described by the legend.}
\label{fig:Z-grad-cc}
\end{figure}

Considering just the CC groups, we can also look for any impact from the
central radio sources on the temperature distribution.
\citet{johnson09b} calculated the temperature gradient from 1T spectral
models, measured from the
temperature peak to the temperature at 0.01\,\RF. This allows a
calculation of the temperature gradient inside the temperature
peak. Performing a Kendall correlation test between the logarithmic 1.4\,GHz
radio power of the BGG and the temperature drop inside the core (normalised
by the mean temperature of the system), we find no significant correlation.
In Figure~\ref{fig:Z-grad-cc} we show the
abundance gradient (calculated as for Figure~\ref{fig:Z-radio}) versus the
temperature gradient in the CC systems. Here we have split the groups
into four bins in the logarithm of the 1.4\,GHz radio power. Looking
at these two parameters in conjunction with the abundance gradient
measurement, steeper metallicity gradients occur in the cores of
the systems with lower radio power, as was suggested by 
Figure~\ref{fig:Z-radio}. Figure~\ref{fig:Z-grad-cc} also confirms that
the temperature gradient shows no strong trend with radio power.

\begin{figure}
\includegraphics[width=8cm]{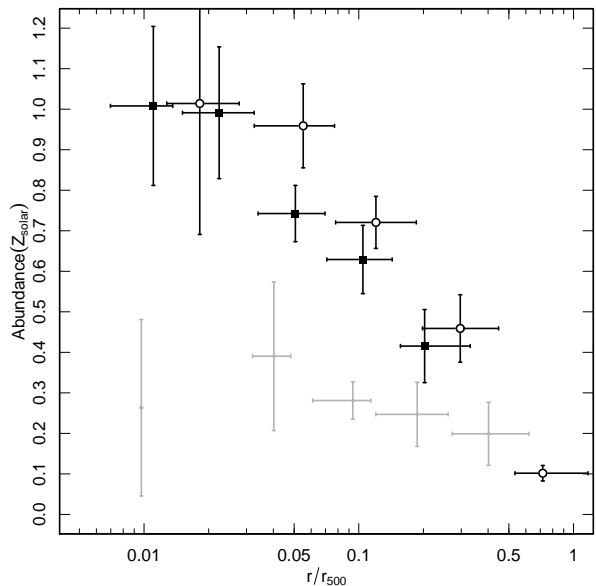}
\caption{The stacked abundance profiles (from the 2T fits) of the 
  CC sample (black),
  calculated from the mean in each radial bin, divided into those
  systems with `radio quiet' BGGs (filled squares) and `radio loud'
  BGGs (open circles). The mean profile of NCC systems from
  Figure~\ref{fig:Z} is shown as grey errorbars. Vertical error bars
  are show the standard error of the mean profile and horizontal error bars show the radial width
  of each bin. }
\label{fig:Z-radio-CC-NCC}
\end{figure}

To further investigate the relationship between AGN activity and
abundance profiles, we sub-divide the 20 groups for which radio
luminosities are available into `radio loud' and `radio quiet'
sub-samples based on the median logarithmic radio luminosity of the
whole sample (log\,\LR\ = 21.96\,W\,Hz$^{-1}$). The number of groups
in each category is shown in Table~\ref{tab:RLRQ}. We show the stacked
abundance profiles for the CC radio loud/radio quiet samples in
Figure~\ref{fig:Z-radio-CC-NCC}, derived from the 2T fits. 
The small number of groups in the
radio loud and radio quiet NCC samples, and the diverse properties of
these groups, preclude us from being able to draw any reliable
conclusions on the effect of a central radio source in NCC systems, so
we show the mean NCC abundance profile from Figure~\ref{fig:Z} in
Figure~\ref{fig:Z-radio-CC-NCC}. The profiles of CC groups with
radio loud and radio quiet BGGs are quite similar. In particular, the
central abundance levels (within 0.05\,\RF) are comparable, and both
subsamples seem to show a relatively flat profile within this radius. 
The large standard errors apparent in Figure~\ref{fig:Z} within the inner
bins reflect considerable group-to-group diversity, as is apparent
from examination of Figure~\ref{fig:Z-CC}. It can be seen that many
CC groups show a plateau or decline in abundance at small radius, whilst
others do not. This diversity makes the formal
significance of the central flattening in abundance marginal within
the stacked data. It is worth recalling that RP07, using higher
spatial resolution \textit{Chandra} data, found a central dip in
abundance in most of their CC-dominated group sample. Dividing our
sample by radio loudness, it can be seen from Figure~\ref{fig:Z} that
there is some indication that the central abundance ``plateau'' in
radio loud BGGs may extend further (to $\sim$0.1\,\RF) than for radio
quiet BGGs (to $\sim$0.05\,\RF). 

\begin{table}
  \centering
  \caption{The breakdown of CC and NCC groups into the `radio quiet'
  and `radio loud' sub-samples.}
  \label{tab:RLRQ}
  \begin{tabular}{ccc}
  \hline
  & CC & NCC\\
  \hline
  Radio quiet & 7 & 3\\
  Radio loud & 7 & 3\\
  \hline
  Total & 14 & 6\\
  \end{tabular}
\end{table}

\subsection{Substructure}

One way of destroying CCs is through mergers, which disrupt the ICM,
mixing the gas. There is observational evidence to support this
picture \citep[e.g.][]{sanderson09}. To test the hypothesis that NCC
groups have had their CCs destroyed by mergers, we can look for
substructure indicative of such events in our group sample.

\begin{figure}
\includegraphics[width=8cm]{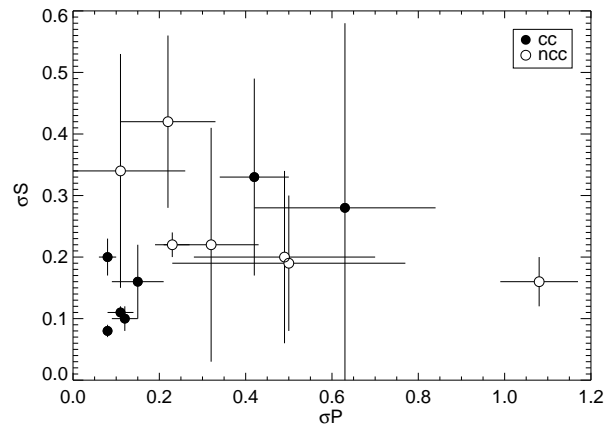}
\caption{The dispersion in entropy versus pressure from the mean
  radial profiles calculated by \citet{finoguenov07} for 14 of the
  groups in our sample. Filled circles show CC systems, and open
  circles show NCC systems. The NCC systems show marginally more
  substructure (see text). }
\label{fig:sub}
\end{figure}

\citet{finoguenov07} measured the level of substructure for 14 of the
systems in our sample, by computing the entropy and pressure
dispersion from the mean radial profiles. 
These gas properties
were derived from 1T spectral fits, but in regions where the
gas exhibits a range of temperatures, the 1T fit gives a 
temperature estimate which is essentially
an emission-weighted mean, and the derived values of pressure and entropy
are also reasonable estimates of the true values within the region.
Figure~\ref{fig:sub} shows
the entropy dispersion versus the pressure dispersion for the 14
groups presented by \citet{finoguenov07}. Examining the mean
dispersion levels in the CC and NCC groups, we find a mean pressure
dispersion of 0.23$\pm$0.08 and 0.42$\pm$0.12 respectively, a
difference that is marginally significant (1.3\,$\sigma$). The mean
dispersion in entropy for CC and NCC groups is 0.18$\pm$0.04 and
0.25$\pm$0.04 respectively, which has similar significance
(1.4\,$\sigma$). The majority (5/7) of the CC groups cluster at low
values of entropy and pressure dispersion. The two outlying CC systems
are HCG\,42 and NGC\,4325.  \citet{finoguenov07} identify NGC\,4325 as
deviating strongly from the mean pressure and entropy profiles, with a
strongly peaked surface brightness profile. HCG\,42 on the other hand
is identified as a `normal' group, albeit with a region of enhanced
pressure and reduced entropy to the east of the group centre
\citep{finoguenov07}.

We are dealing with less than half the group sample here, but there is
tentative evidence to suggest a higher level of dispersion in both
pressure and entropy in NCC systems. This indicates that the ICM in
these systems shows more substructure. This might result from recent
large scale disturbance of the gas in NCC groups, such as could be
produced by a merger or possibly a major AGN outburst. However, we
note that the NCC groups examined here are typically cooler than the
CC groups, and so should have shallower group potential wells. Hence
the `lumpiness' due to the individual galaxy members will be greater
in the NCC systems, and this may in turn contribute to the scatter
seen in entropy and pressure.  Applying a Kendall rank coefficient
correlation test to test for a correlation between the group
temperature and the entropy dispersion shows no significant
correlation; however, performing the same test on group temperature
and pressure dispersion, we find a highly significant anticorrelation
-- cooler groups tend to have higher pressure dispersion. The
correlation test returns a value for $\tau$ of $-0.57$ and a p-value
of $< 1$~per~cent. The pressure may therefore be responding to the
increased `lumpiness' of the group potential in lower temperature
systems.

A further avenue for exploring substructure is to consider the motion
of the BGG relative to the rest of the group. Central dominant
galaxies in clusters with evidence for substructure tend to have large
peculiar velocities \citep{bird94}. To probe the dynamical disturbance
of CC and NCC groups, we calculated the absolute difference between
the velocity of the BGG and the mean velocity of the remaining group
members. In CC groups, we find a mean velocity offset of $136\pm
26$~km~s$^{-1}$, whereas for NCC groups we find a mean velocity offset
of $218\pm 80$~km~s$^{-1}$, where the errors quoted are the standard
errors on the mean velocity offsets. Typically, brightest cluster galaxies have
a peculiar velocity of approximately one third the velocity dispersion
of their host cluster \citep{coziol09}. The typical velocity
dispersion of our group sample is $\sim 400$~km~s$^{-1}$, making the
velocity offset seen in CC groups consistent with this observation.
However, NCC groups show a higher velocity offset, which may again
indicate a higher level of dynamical disturbance in comparison to CC
groups.

\section{Discussion}
\label{sec:discuss}
There is a clear difference in the stacked abundance profile
properties of CC and NCC galaxy groups. Considering the linear fits to
the raw data points (shown in Figure~\ref{fig:Z}), we find the NCC
systems to show approximately flat abundance profiles, compared to the
steep gradients observed in the CC systems. Excluding the group
HCG\,51, which has abnormally high abundance at large radii in the 1T
fits, increases the abundance gradient, but this remains much shallower 
than that observed in the CC systems, and remains consistent with 
a flat profile in the 2T fits. The abundance profiles of NCC groups also
contrast sharply with those of NCC clusters, as shown in
Figure~\ref{fig:clusters}.

All groups might be expected to possess cool cores, given their short
central cooling times -- typically a few Gyr at 0.05\,\RF\ (O'Sullivan
et al, in prep). One would also expect metallicity gradients in
undisturbed groups; given the luminosities of their BGGs, type Ia
supernovae products should be available to build the central metal
peak on a timescale of $\sim$\,5\,Gyr \citep{rasmussen09}. The
difference in the abundance profiles between CC and NCC groups, most
notably the lack of a central peak in NCC groups, therefore suggests
that gas in NCC groups has been vigorously mixed within the past few
Gyr.

\begin{figure}
\includegraphics[width=8cm]{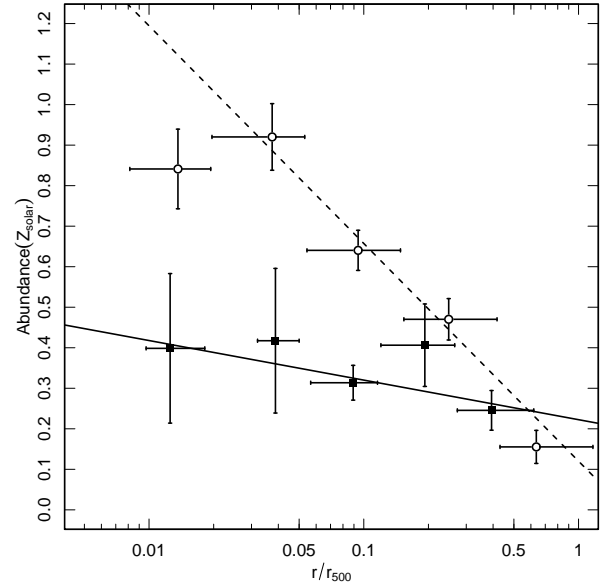}
\caption{The stacked abundance profiles of the CC (open circles) and
  NCC sub-samples (filled squares) chosen to have the same mean
  temperatures. The vertical error bars show the standard error on the
  mean,
  and the horizontal error bars show the radial width of each bin. The
  dashed and solid lines are the linear fits to the whole data set of
  abundances shown in Figure~\ref{fig:Z}.}
\label{fig:Z:kT}
\end{figure}

\citet{rasmussen09} found evidence that galaxies in low-mass ($T\la
1$~keV) groups are less efficient at releasing metals than those in
more massive groups.  This could help to explain the lower central
abundances seen in NCC systems, since the NCC groups in our sample are
typically cooler than the CCs, although there is a substantial overlap
in the temperatures of the two sub-samples \citep{johnson09b}. To
remove any systematic effect associated with temperature, we removed
the hottest and coolest groups from our sample, leaving sub-samples of
13 CC groups and 8 NCC groups spanning the same temperature
range. These two trimmed sub-samples have very similar mean
temperatures (0.98$\pm$0.06\,\keV\ for CCs, and 0.96$\pm$0.08\,\keV\
for NCCs). The stacked abundance profiles of these trimmed sub-samples
are shown in Figure~\ref{fig:Z:kT}, and are very similar to those for
the full CC and NCC samples, shown in Figure~\ref{fig:Z}, although the
innermost point for this subset of CC groups has dropped by approximately
0.13$\Zsol$.

We next examine whether the differences in metallicity could result
from differences in star formation efficiency of CC and NCC groups, by
determining the ratio of the group $B$-band luminosity to total group
mass. For self-similar systems, total mass $M$ scales as $T^{3/2}$, so
we use the latter as a proxy for mass. We find $L_{B,group}/T^{3/2} =
(1.57\pm 0.18)\times 10^{11}$~L$_{\odot,B}$~keV$^{-3/2}$ for the CC
groups and $L_{B,group}/T^{3/2} = (2.02\pm 0.45)\times
10^{11}$~L$_{\odot,B}$~keV$^{-3/2}$ for the NCC groups. These values
agree within their quoted standard errors, suggesting that CC and NCC
systems are similarly efficient at forming stars. We can therefore
exclude the possibility that NCC groups produce significantly fewer
metals per unit total mass.

Examining the gas mass weighted metal fraction in CC and NCC systems
showed that if the central peak in NCC groups has been mixed out, we
would expect an increase in metallicity at large radius in NCCs
compared to CCs. There is no evidence for this given the limits of our
data, implying that if the central peak has been mixed out, the metals
have been pushed out past 0.5\,\RF. To fully resolve this issue would
require detailed metallicity measurements at radii beyond 0.5\,\RF.

Taking an Occam's Razor approach, we seek to identify a single cause
for both the lack of cool cores and the flatter abundance profiles in
NCC groups. Three possible mechanisms might account for the lack of
cool cores in some groups -- pre-heating, AGN mixing and merger
disruption. We will now explore these three in detail, to see how they
compare with observations. Feedback from supernova driven winds
provides a fourth possible mechanism for suppressing cooling in the
centres of groups. However, a higher level of such feedback would lead
to an increased metal mass fraction, which is not observed, pointing
instead to AGN feedback as the dominant feedback mechanism
\citep{johnson09b}. This conclusion was also reached by
\citet{diehl08b}, who found that although supernovae feedback can play
a part in balancing radiative cooling in X-ray faint early-type
galaxies, it is not the dominant mechanism in X-ray bright systems.

\subsection{Pre-heating}
One mechanism capable of preventing the formation of a cool core is
strong initial pre-heating, which raises the entropy of the system to
a level that prevents its cooling by the present day. This was
proposed as a framework for understanding CC and NCC behaviour by
\citet{mccarthy08}. However, if NCC groups were formed via pre-heating
the system prior to group collapse as envisaged by the
\citet{mccarthy08} model, there is no reason why abundance gradients
could not have built up since, as the estimated typical enrichment
time required to build the central iron peak is $\sim$5\,Gyr
\citep{rasmussen09}. This method is also inconsistent with the short
central cooling times in NCC groups (O'Sullivan et al, in prep). It
seems that we can therefore rule out pre-heating as a means capable,
on its own, of preventing the formation of a CC, whilst accounting for
the flat abundance gradients in NCC groups.

\subsection{AGN mixing}
AGN activity can re-distribute the metals in the ICM; for example, the
entrainment of enriched gas has been observed in the Perseus
\citep{sanders05} and Hydra A clusters \citep{kirkpatrick09}. In our
sample, we see central flattening (or dips) in the abundance profiles of many CC
groups, and there is some evidence that this
flat region may extend further (to $\sim$0.1\,\RF) in the CC groups
with a  `radio-loud' (\LR\ $> 21.96$~W~Hz$^{-1}$) BGG, compared to an
extent of $\sim$0.05\,\RF\ around radio quiet BGGs. The abundance declines
beyond these radii, suggesting that the effect of significant ongoing
AGN activity in CC systems is to mix the ICM within the central 
$\sim$50\,kpc. These observations fit
well with the AGN-driven circulation flow model of \citet{mathews04},
which produces a central positive temperature gradient and a flat core
in the iron abundance within $\sim 50$~kpc.

One potential difficulty with mixing gas out to 0.1\,\RF\ in
radio-loud CC groups is the required timescale. If the advection of
enriched gas is restricted to velocities below the sound speed ($\sim
500$~km~s$^{-1}$ for a $T=1$~keV system) then such mixing would take
approximately 10$^{8}$ years.  This is longer than most estimates of
the duration of AGN outbursts.  However, powerful FR~II outbursts
expand at supersonic velocities, generating lobes with typical size
scales of 100~kpc \citep[e.g.][]{shabala08}. 
With the current data we
cannot probe the relationship between the current radio activity of
the central BGG in NCC systems and their abundance profiles. However,
the results of \citet{diehl08b}, who found a lack of cool cores in
elliptical galaxies with low radio luminosity, show that this is an
important area for future study.

\subsection{Mergers}
\citet{sanderson06} find the NCC clusters in their sample to show
evidence of recent disruption, favouring a merger hypothesis for their
formation. Furthermore, \citet{sanderson09} have shown that weaker 
central cooling correlates with increased dynamical disturbance in clusters.  
Merging occurring on the group scale could therefore
disrupt a cool core, and could potentially mix the metal
distribution. We do find evidence for a significantly larger degree of
substructure in the ICM of NCC groups, which could result from the
impact of recent merger activity on the ICM. However, given the
typically lower temperatures of NCC systems, the observation of
greater entropy and pressure dispersion could alternatively be a
reflection of the response of the ICM to the presence of individual
galaxies. On the other hand, enhanced substructure could be an
indicator of mixing processes, which need not necessarily be caused by
mergers. With the current data there is no \textit{unambiguous} sign
of enhanced merger activity in NCC groups, but this is a promising
area for further study.

On the basis of numerical simulations, \citet{burns08} propose that
the early merger history of NCC clusters establishes their gas
properties, as major mergers destroy cool cores, leading to hot cores
with long cooling times. However, this model fails in much the same
way as the pre-heating models, as the observed short central cooling
times in NCC groups and clusters \citep[e.g.][O'Sullivan et al, in
prep.]{sanderson08} and the lack of abundance gradients in NCC groups
cannot be explained.

Simulations of merging galaxy clusters that initially host metallicity
gradients by \citet{poole08} failed to mix the ICM sufficiently to
wipe out the metallicity gradient. However, the presence of a high
abundance region in HCG\,62, approximately 34\,\kpc\ from the peak of
the group X-ray emission provides observational evidence to support
merger-induced metal mixing \citep{gu07}. It is true that \citet{gu07}
cannot rule out AGN-induced mixing, and given the strong cool core in
this system \citep{johnson09b}, the merging activity appears not to
have disrupted the cool core, or it occurred sufficiently long ago to
allow the subsequent formation of a CC. It is possible that
pre-heating could work in conjunction with mergers in NCCs to both
prevent cooling and to mix the distribution of metals (I.~G.~McCarthy,
private communication).

An alternative to \textit{destruction} of a central abundance peak
might be models in which the peak has not been established in the
first place.  Since much of the central peak is believed to originate
from the BGG, this can be achieved if the BGG has not long been
located at the centre of the group. This could be the case if NCC
groups are dynamically young, and have only recently collapsed. This
could also account for the presence of substructure in these systems.
Even in older groups, it is worth noting that one consequence of a
recent merger would be to displace the BGG from the group
centre. Until the post-merger system settles down into a new
virialised state, any new enriched gas from the galaxy will be
distributed over a larger region than would be the case in an
undisturbed group. The larger velocity offset of the BGG relative to
the rest of the group seen in NCC groups compared to CC groups is
consistent with this idea.

\section{Conclusions}
\label{sec:conclude}
Using an \XMM\ sample of 28 galaxy groups presented by
\citet{johnson09b}, we have determined the mean abundance profiles of
cool core (CC) and non cool core (NCC) systems. This is the first time
that the abundance profiles of NCC groups have been examined in
detail. Fitting a linear model in log--linear space to abundance 
profiles derived from two--temperature spectral fits, we find CC groups
to exhibit steep abundance gradients with a slope of $-0.54\pm
0.07$. We find generally good agreement between the abundance profiles of
the CC groups and the abundance profiles of the {\em Chandra} group
sample of \citet{rasmussen07}, which was dominated by CC
systems. However, fitting a linear model to the profiles of NCC groups
yields a result consistent with a flat profile.

Examining the gas mass weighted mean metal fraction, the consequence
of mixing out a central peak in the abundance profile of a NCC group
would be a significant increase in metallicity compared to CC systems
at large radii. However, there is no evidence for such an enhanced
metallicity out to 0.5\,\RF, the limit of our data. To ascertain
whether a central peak is simply mixed out to large radius in NCC
systems requires mapping the metallicity of NCC groups at radii
greater than 0.5\,\RF.

We investigate current AGN activity through the 1.4\,GHz radio power
of the brightest group galaxies (BGGs) in the groups, and find a
significant anticorrelation between the core abundance gradient and the radio
power of the BGG within CC systems. 
Dividing the CC groups into `radio loud' (log\,\LR\ $>$ 21.96\,W\,Hz$^{-1}$
 and `radio quiet' (log\,\LR\ $\leq$ 21.96\,W\,Hz$^{-1}$)
subsamples shows both to have broadly similar abundance profiles. 
Central flattening in the typical abundance profiles may extend to larger radius 
($\sim$0.1\,\RF) in radio-loud systems compared
to radio-quiet systems ($\sim$0.05\,\RF). We interpret this as
tentative evidence for AGN mixing within the central regions. Given the 
limits of the data, it is not possible to
probe the behaviour of NCC abundance profiles in terms of the current
radio activity of their BGGs.

The lack of cool cores and evidence for enhanced substructure in NCC
groups suggests a disruptive event, whilst the short central cooling
times and timescales for building a central metal peak indicate that
this disruption has occurred within the past few Gyr. We favour merging as a
likely mechanism consistent with the observations, although we
cannot rule out powerful AGN outbursts as the drivers for group-wide
mixing in NCC groups. The viability of powerful AGN outbursts in accounting for
the properties of NCC groups could be tested by large statistical
studies. It is also possible that some NCC groups are young systems,
which have recently collapsed for the first time. Given the diverse
properties of NCC groups, it is quite possible that they do not form a
homogeneous class, and that a number of different effects are at work.

Critical questions for further investigation of NCC groups are to
establish to what extent the presence of a currently active AGN
affects their metallicity profiles, and whether enriched gas has been
boosted to high entropy, by either mergers or AGN activity, and hence
relocated to large radii in these groups.

\section*{Acknowledgments}
We thank the referee for their helpful comments, which have improved the 
paper. We thank Ian McCarthy for interesting discussions in
relation to this project. RJ acknowledges support from STFC/PPARC and
the University of Birmingham. JR acknowledges support by the 
Carlsberg Foundation. AF acknowledges support from BMBF/DLR under
grants 50OR0207 and 50OR0204 to MPE. This research has made use of the
NASA/IPAC Extragalactic Database (NED) which is operated by the Jet
Propulsion Laboratory, California Institute of Technology, under
contract with the National Aeronautics and Space Administration.



\label{lastpage}

\end{document}